\documentclass[11pt,twocolumn,a4paper]{article}

\usepackage{amsmath}
\usepackage{times}
\usepackage{mathptmx}
\usepackage{natbib}
\usepackage{graphicx}
\usepackage{psfig}
\usepackage[table]{xcolor}
\usepackage{ltxtable}
\usepackage{multirow}
\usepackage{epsfig}

\newcommand{\Mo}{$\mbox{M}_\odot$}

\begin{document} 
\title{Autonomous Spacecraft Navigation With Pulsars}
\author{Werner Becker$^{1,2}$\thanks{wbecker@mpe.mpg.de},\, Mike G.~Bernhardt$^{1}$,\, Axel Jessner$^{2}$
       \\ \\ \small{$^1$Max-Planck-Institut f\"ur extraterrestrische Physik, Gie\ss{}enbachstra\ss{}e, 85748 Garching, Germany}
       \\    \small{$^2$Max-Planck-Institut f\"ur Radioastronomie, Auf dem H\"ugel 69, 53121 Bonn, Germany}
}
\date{March 13, 2013}

\label{paper:WBecker}
\maketitle

\begin{abstract}
 An external reference system suitable for deep space navigation can be defined by
 fast spinning and strongly magnetized neutron stars, called pulsars. Their beamed 
 periodic signals have timing stabilities comparable to atomic clocks and provide 
 characteristic temporal signatures that can be used as natural navigation beacons, 
 quite similar to the use of GPS satellites for navigation on Earth. By comparing 
 pulse arrival times measured on-board a spacecraft with predicted pulse arrivals 
 at a reference location, the spacecraft position can be determined autonomously 
 and with high accuracy everywhere in the solar system and beyond. The unique 
 properties of pulsars make clear already today that such a navigation system 
 will have its application in future astronautics. In this paper we describe
 the basic principle of spacecraft navigation using pulsars and report on the 
 current development status of this novel technology.
\end{abstract}

\section{Introduction}
 Today, the standard method of navigation for interplanetary spacecraft is a 
 combined use of radio data, obtained by tracking stations on Earth, and optical 
 data from an on-board camera during encounters with solar system bodies. Radio 
 measurements taken by ground stations provide very accurate information on the 
 distance and the radial velocity of the spacecraft with typical random errors of 
 about 1\,m and 0.1\,mm/s, respectively \citep{madde2006}. The components of 
 position and velocity perpendicular to the Earth-spacecraft line, however, are 
 subject to much larger errors due to the limited angular resolution of the radio 
 antennas. Interferometric methods can improve the angular resolution to about 
 25\,nrad, corresponding to an uncertainty in the spacecraft position of about 
 4\,km per astronomical unit (AU) of distance between Earth and spacecraft 
 \citep{james2009}. With increasing distance from Earth, the position error 
 increases as well, e.g., reaching a level of uncertainty of the order of 
 $\pm 200$\;km at the orbit of Pluto and $\pm 500$\;km at the distance of Voyager~1. 
 Nevertheless, this technique has been used successfully to send space probes to 
 all planets in the solar system and to study asteroids and comets at close range. 
 However, it might be necessary for future missions to overcome the disadvantages 
 of this method, namely the dependency on ground-based control and maintenance, 
 the increasing position and velocity uncertainty with increasing distance 
 from Earth as well as the large propagation delay and weakening of the signals 
 at large distances. It is therefore desirable to automate the procedures of orbit 
 determination and orbit control in order to support autonomous space missions.

 Possible implementations of autonomous navigation systems were already discussed 
 in the early days of space flight \citep{battin1964}. In principle, the orbit of a
 spacecraft can be determined by measuring angles between solar system bodies and 
 astronomical objects; e.g., the angles between the Sun and two distant stars and a 
 third angle between the Sun and a planet. However, because of the limited angular 
 resolution of on-board star trackers and sun sensors, this method yields spacecraft 
 positions with uncertainties that accumulate typically to several thousand kilometers. 
 Alternatively, the navigation fix can be established by observing 
 multiple solar system bodies: It is possible to triangulate the spacecraft position
 from images of asteroids taken against a background field of distant stars. This 
 method was realized and flight-tested on NASA's Deep-Space-1 mission between 
 October 1998 and December 2001. The Autonomous Optical Navigation (AutoNav) system 
 on-board Deep Space~1 provided the spacecraft orbit with $1\sigma$ errors of 
 $\pm 250$\,km and $\pm 0.2$\,m/s, respectively \citep{riedel2000}. 
 Although AutoNav was operating within its validation requirements, 
 the resulting errors were relatively large compared to ground-based navigation.

 In the 1980s, scientists at NRL (United States Naval Research Laboratory)
 proposed to fly a demonstration experiment called the Unconventional
 Stellar Aspect (USA) experiment \citep{wood1993}. Launched in 1999 on the
 Advanced Research and Global Observation Satellite (ARGOS), this experiment
 demonstrated a method of position determination based on stellar occultation
 by the Earth's limb as measured in X-rays. This technique, though, is limited 
 to satellites in low Earth orbit. 

 An alternative and very appealing approach to autonomous spacecraft navigation 
 is based on pulsar timing. The idea of using these celestial sources as a 
 natural aid to navigation goes back to the 1970s when \citet{downs1974} investigated 
 the idea of using pulsating radio sources for interplanetary navigation. 
 \citet{downs1974} analyzed a method of position determination by comparing pulse 
 arrival times at the spacecraft with those at a reference location. Within the 
 limitations of technology and pulsar data available at that time (a set of
 only 27 radio pulsars were considered), Downs showed that spacecraft position errors 
 on the order of 1500\,km could be obtained after 24 hours of signal integration. 
 A possible improvement in precision by a factor of 10 was estimated if better 
 (high-gain) radio antennas were available for the observations. 

 \citet{chester+butman1981} adopted this idea and proposed to use X-ray pulsars, of 
 which about one dozen were known at the time, instead of radio pulsars. They estimated 
 that 24 hours of data collection from a small on-board X-ray detector with
 0.1\,m$^2$ collecting area would yield a three-dimensional position accurate to 
 about 150\,km. Their analysis, though, was not based on simulations or actual pulsar 
 timing analyses; neither did it take into account the technological requirements or 
 weight and power constraints for implementing such a navigation system. 

 These early studies on pulsar-based navigation estimated relatively large position 
 and velocity errors so that this method was not considered to be an applicable 
 alternative to the standard navigation schemes. However, pulsar astronomy has improved 
 considerably over the last 30 years since these early proposals. Meanwhile, pulsars 
 have been detected across the electromagnetic spectrum and their emission properties 
 have been studied in great detail \citep[cf.][for a collection of comprehensive reviews 
 on pulsar research]{beckerBook2009}. Along with the recent advances in detector and 
 telescope technology this motivates a general reconsideration of the feasibility and 
 performance of pulsar-based navigation systems. The present paper reports on our latest
 results and ongoing projects in this field of research. Its structure is as follows: 
 After summarizing the most relevant facts on pulsars and discussing which pulsars are  
 best suited for navigation purposes in \S\,\ref{TypesOfPulsars}, we briefly describe 
 the principles of pulsar-based navigation in \S\,\ref{PrinciplesPulsarNavigation}. 
 Pulsars emit broadband electromagnetic radiation which allows an optimization for the 
 best suited waveband according to the highest number of bits per telescope collecting 
 area, power consumption, navigator weight and compactness. A possible antenna type and 
 size of a navigator which detects pulsar signals at 21\,cm is described in 
 \S\,\ref{RadioAntenna}. In \S\,\ref{XNAV} we discuss the possibility of using X-ray  
 signals from pulsars for navigation. The recent developments of low-mass X-ray mirrors 
 and active-pixel detectors, briefly summarized in \S\,\ref{XNAV_hardware}, makes it very 
 appealing to use this energy band for pulsar-based navigation. 

\section{The Various Types of Pulsars and their Relevance for Navigation\label{TypesOfPulsars}}

 Stars are stable as long as the outward-directed thermal pressure, caused by nuclear 
 fusion processes in the central region of the star, and the inward-directed gravitational 
 pressure are in equilibrium. The outcome of stellar evolution, though, depends solely on 
 the mass of the progenitor star. A star like our sun develops into a white dwarf. Stars
 above $\approx 8$ \Mo\ undergo a gravitational collapse once their nuclear fuel is depleted. 
 Very massive stars of more than about 30 \Mo\ end up as black holes and stars in the 
 intermediate mass range of about 8 to 30 \Mo\ form neutron stars. It is assumed that a 
 neutron star is the result of a supernova explosion, during which the bulk of its 
 progenitor star is expelled into the interstellar medium. The remaining stellar core 
 collapses under its own weight to become a very compact object, primarily composed of 
 neutrons -- a neutron star. With a mass of typically 1.4 \Mo, compressed into a sphere 
 of only 10\,km in radius, they are quasi gigantic atomic nuclei in the universe. Because 
 of their unique properties they are studied intensively by physicists of various disciplines 
 since their discovery in 1967 \citep{hewish1968}.

\begin{figure}[b!!!!!!!!!!]
  \centerline{\psfig{file=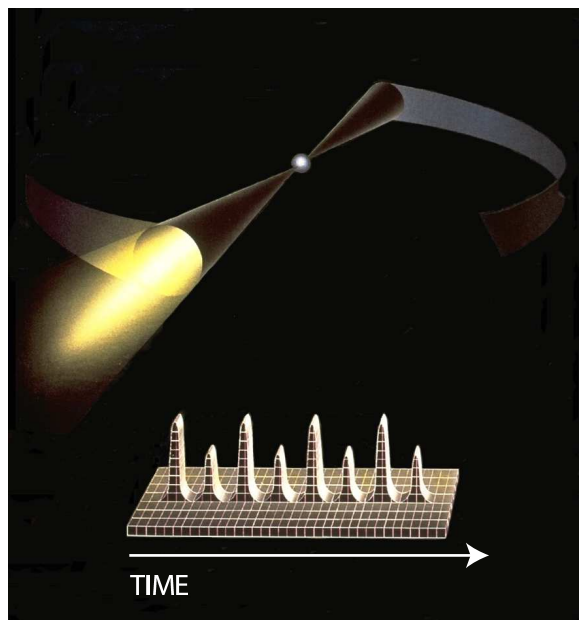,width=7.5cm,clip=}}
  \caption{\small Artist's impression of a rotation-powered pulsar. The neutron star 
   appears as a pulsating source of radiation if the rotating emission beam crosses 
   the observer's line of sight. Averaging these periodic pulses of intensity over 
   many rotation cycles results in a stable pulse profile. Because of the timing 
   stability of most pulsars, the arrival time of pulses can be predicted with very 
   high precision, which is an essential requirement for a navigation system based on 
   pulsar observations.} \label{image:RotPowPSR}
  \end{figure}

 Fast spinning and strongly magnetized neutron stars are observable as pulsars if their spin 
 axis and magnetic field axis are not aligned. Having co-rotating magnetic fields of 
 $B_\perp \approx 10^9$--$10^{13}$\,G and spin periods down to milliseconds they radiate 
 broadband electromagnetic radiation along narrow emission cones. If this radiation cone 
 crosses the observer's line of sight a pulse of intensity is recorded in the observing 
 device (cf.~Figure~\ref{image:RotPowPSR}). 
 The name {\em Pulsar} refers to this property. They have been discovered by their radio signals
 \citep{hewish1968}. In source catalogs their common abbreviation is therefore {\em PSR} which 
 stands for Pulsating Source of Radio, although they have also been detected in other bands 
 of the electromagnetic spectrum meanwhile. Three different classes of pulsars can be distinguished 
 according to the energy source of their electromagnetic radiation. As we will see, only one class 
 is suitable for spacecraft navigation:

\begin{itemize}
  \item {\bf Accretion-powered pulsars} are close binary systems in which a
  neutron star is accreting matter from a companion star, thereby gaining
  energy and angular momentum. There are no radio waves emitted from the accretion 
  process, but these systems are bright in X-rays. The observed X-ray
  pulses are due to the changing viewing angle of a million degree hot spot 
  on the surface of the neutron star. These hot spots
  are heated by in-spiraling matter from an accretion disk. The accretion 
  disk and the accretion column itself can also be sources of X-rays. The spin 
  behavior of accretion-powered pulsars can be very complicated and complex. They 
  often show an unpredictable evolution of rotation period, with erratic changes 
  between spin-up and spin-down as well as X-ray burst activities \citep{ghosh2007}. 
  Although accretion-powered pulsars are usually bright X-ray sources, and thus 
  would give only mild constraints on the sensitivity requirements of a 
  pulsar-based navigation system, their unsteady and non-coherent timing behavior 
  disqualifies them as reference sources for navigation.

 \item {\bf Magnetars} are isolated neutron stars with exceptionally high magnetic 
  dipole fields of up to $10^{15}$\,G. All magnetars are found to have rotation periods 
  in the range of about 5 to 10 seconds. RXTE and other X-ray observatories have detected 
  super-strong X-ray bursts with underlying pulsed emission from these objects. According 
  to the magnetar model of \citet{duncan1992}, their steady X-ray emission is powered by 
  the decay of the ultra-strong magnetic field. This model also explains the X-ray burst 
  activity observed from these objects, but there are also alternative theories, which
  relate these bursts to a residual fall-back disk \citep{truemper2010, truemper2013}.  
  However, their long-term timing behavior is virtually unknown, which invalidates 
  these sources for the use in a pulsar-based spacecraft navigation system. 
\end{itemize}

\noindent
  Concerning their application for navigation, the only pulsar class that really 
  qualifies is that of rotation-powered ones. 

\begin{itemize}
 \item {\bf Rotation-powered pulsars} radiate broadband electromagnetic radiation (from radio 
  to optical, X- and gamma-rays) at the expense of their rotational energy, i.e., the pulsar 
  spins down as rotational energy is radiated away by its co-rotating magnetic field. The amount 
  of energy that is stored in the rotation of the star can be estimated as follows: A neutron 
  star with a radius of $R=10$ km and a mass of $M=1.4$~\Mo\ has a moment of inertia  $I\approx 
  (2/5)\,M\, R^2 \approx 10^{45}\,\mbox{g cm}^2$. The rotational energy of such a star is 
  $E_\text{rot}=2 \pi^2\, I\, P^{-2}$. Taking the pulsar in the Crab nebula with $P\approx 33$\,ms 
  as an example, its rotational Energy is $E_\text{rot} \approx 2\times 10^{49}\,\mbox{erg}$, 
  which is comparable with the energy released by thermonuclear burning of our sun in hundred 
  million years. The spin period of a rotation-powered pulsar increases with time due to a braking
  torque exerted on the pulsar by its  magneto-dipole radiation. For the Crab pulsar, the observed  
  period derivative is $\dot{P}= 4.2\times 10^{-13}$\,\mbox{s s}$^{-1}$, which implies a decrease 
  in rotational energy of $\dot{E}_\text{rot} = - 4\pi^2 I\, \dot{P}\,P^{-3} \approx 4.5 \times 10^{38}\,
  \mbox{erg s}^{-1}$. It has been found, though, that the spin-down energy is not distributed 
  homogeneously over the electromagnetic spectrum. In fact, only a fraction of about 
  $(10^{-7}-10^{-5})\,\dot{E}_\text{rot}$ is observed in the radio band whereas it is roughly
  $(10^{-4}-10^{-3})\,\dot{E}_\text{rot}$ in the X-ray band and $(10^{-2}-10^{-1})\,\dot{E}_\text{rot}$  
  in the gamma-ray band \citep{becker2009}.

  There are two types of rotation-powered pulsars: (1) Field pulsars have  periods between tens of 
  milliseconds to several seconds and constitute more than 90\% of the total pulsar population. (2) 
  About 10\% of the known pulsars are so-called millisecond pulsars, which are defined to have 
  periods below 20 milliseconds. They are much older than normal pulsars, posses weaker magnetic 
  fields and, therefore, relatively low spin-down rates.  Accordingly, they exhibit very high 
  timing stabilities, which are comparable to atomic clocks \citep{taylor1991,matsakis1997}. This 
  property of millisecond pulsars is of major importance for their use in a pulsar-based navigation 
  system. Figure~\ref{image:p_pdot} clearly shows 
 \begin{figure}[b!!!!!!!!!!!!!]
  \hspace{-0.5ex}{\includegraphics[width=7.75cm, angle=0,clip=]{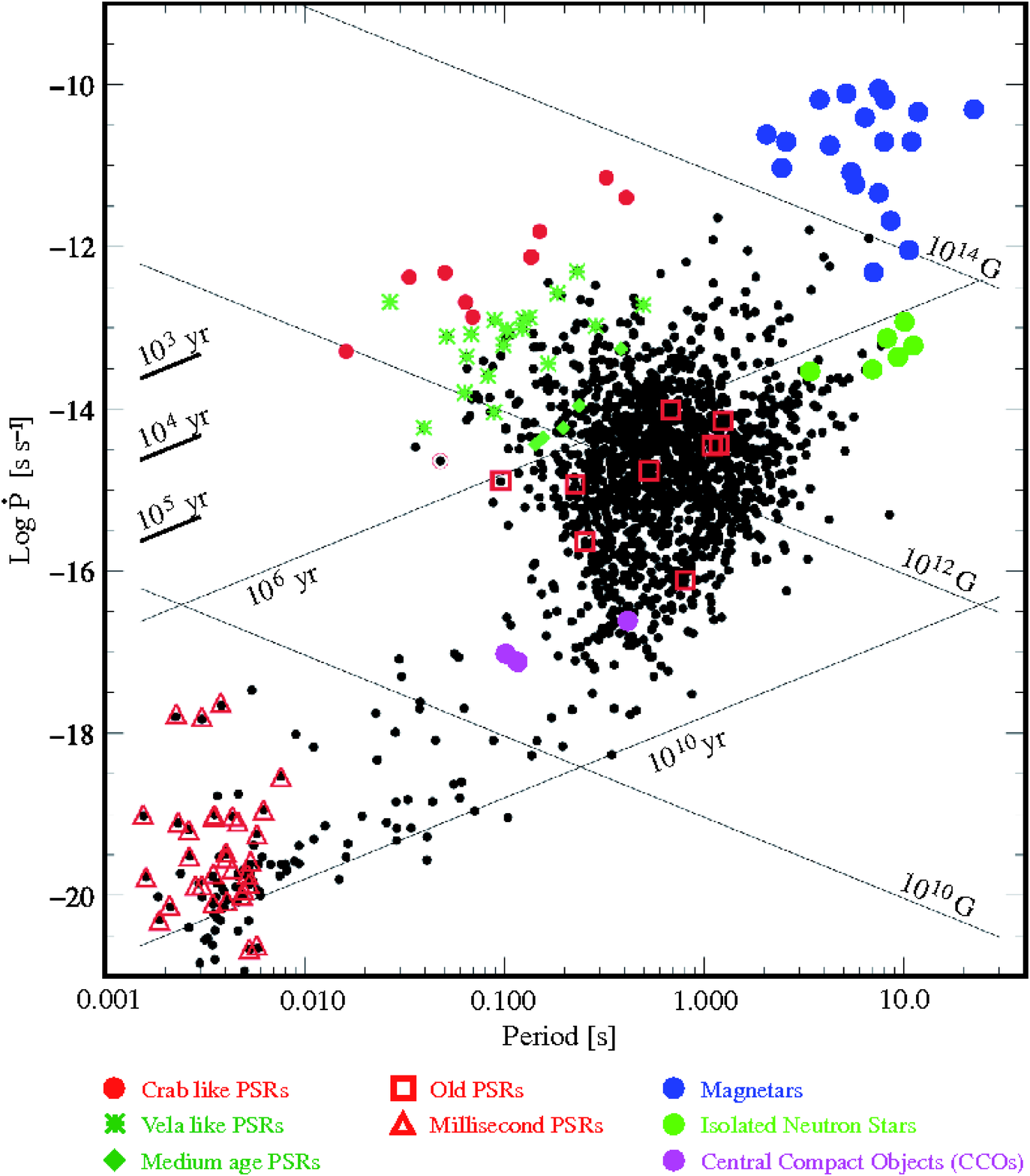}}
  \caption{\small The $P$-$\dot{P}$ diagram; distribution of rotation-powered pulsars according 
   to their spin parameters. X-ray detected pulsars are indicated by colored symbols. The straight 
   lines correspond to constant ages $\tau=P/(2\dot{P})$ and magnetic field strengths 
   $B_\perp=3.2\times 10^{19}(P\dot P)^{1/2}$ as deduced within the framework of the magnetic 
   braking model. 
   }\label{image:p_pdot} 
 \end{figure}
  that these two types of pulsars belong to distinct populations. Most likely, they are connected 
  by an evolutionary process: It is assumed that millisecond pulsars are born as normal pulsars in 
  a close binary system, but their rotation accelerates as they pass through a phase of accretion 
  in which mass and angular momentum are transferred from the evolving companion star to the pulsar 
  \citep[e.g.,][]{bhattacharya1991}. However, the fact that millisecond pulsars are often found in 
  binary systems does not affect their suitability for spacecraft navigation as the binary motion 
  can easily be accounted for in pulsar timing \citep{blandford1976}. Millisecond pulsars -- also 
  referred to as recycled pulsars -- were discovered by \citet{backer1982} and studied 
  extensively in the radio band by, e.g., \citet{kramer1998}. Pulsed X-ray emission from millisecond
  pulsars was discovered by \citet{becker+truemper1993} using ROSAT. However, only XMM-Newton and Chandra 
  had the sensitivity to study their X-ray emission properties in the $0.5-10$ keV  band in greater 
  detail. The quality of data from millisecond pulsars available in the X-ray data archives, though, 
  is still very inhomogeneous. While from several of them high  quality spectral, temporal and spatial 
  information is available, many others, especially those located in globular clusters, are detected 
  with just a handful of events, not allowing, e.g., to constrain their timing and spectral 
  properties in greater detail. From those millisecond pulsars  
  detected with a high signal-to-noise ratio strong evidence is found for a dichotomy of their X-ray 
  emission properties. Millisecond pulsars having a spin-down energy of $\dot{E}\ge10^{35}$\,erg/s 
  (e.g., PSR J0218$+$4232, B1821$-$24 and B1937$+$21) show X-ray emission dominated by non-thermal 
  radiation processes. Their pulse profiles show narrow peaks and pulsed fractions close to   
  100\% (cf.~Figure~\ref{image:pulseprofiles}).  Common for these pulsars is that they show relatively 
  hard X-ray emission, making it possible to study some of them even with RXTE. For example, emission 
  from B1821$-$24 in the globular cluster M28 is detected by RXTE up to $\approx 20$\,keV, albeit with 
  limited photon statistics. For the remaining millisecond pulsars the X-ray emission is found to be much
  softer, and pulse profiles are more sinusoidal.  Their typical fraction of pulsed X-ray
  photons is between 30 and 60\%.

  \begin{figure}[t]   
  \centerline{\includegraphics[width=7.5cm, height=12.8cm,angle=0,clip=]{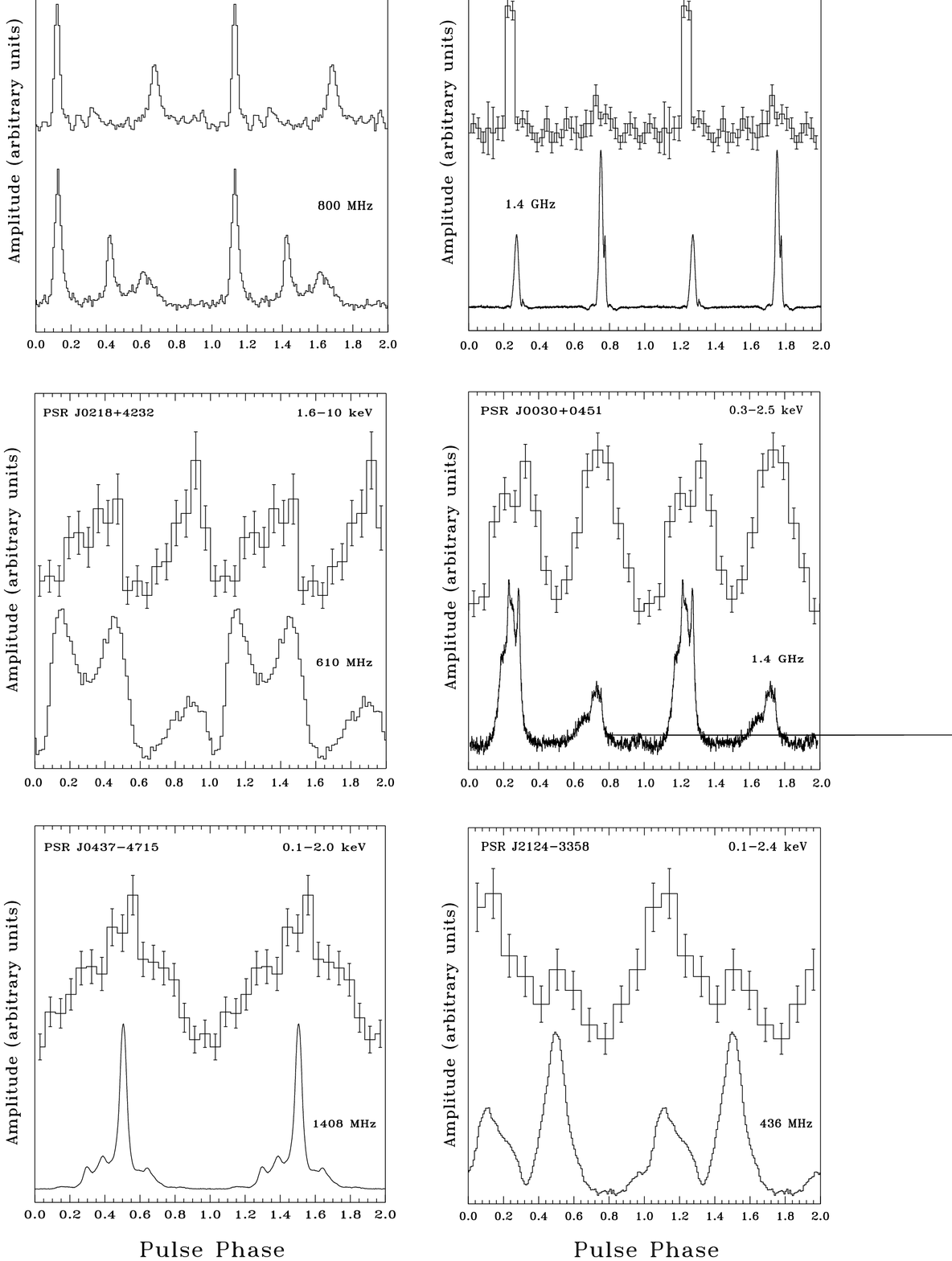}}
   \caption[]{\small X-ray and radio pulse profiles for the six brightest millisecond 
   pulsars. Two full pulse cycles are shown for clarity. From \citet{becker2009}.
   \label{image:pulseprofiles} }
  \end{figure}

\end{itemize}
  
\noindent
  Some rotation-powered pulsars have shown glitches in their spin-down behavior, i.e., abrupt
  increases of rotation frequency, often followed by an exponential relaxation toward the 
  pre-glitch frequency \citep{espinoza2011, yu2013}. This is often observed in young pulsars 
  but very rarely in old and millisecond pulsars. Nevertheless, the glitch behavior of pulsars 
  should be taken into account by a pulsar-based navigation system.

  Today, about 2200 rotation-powered pulsars are known \citep{atnf}. About 150 have been detected 
  in the X-ray band \citep{becker2009}, and approximately 1/3 of them are millisecond pulsars.
  In the past $30-40$ years many of them have been regularly timed with high precision especially 
  in radio observations. Consequently, their ephemerides (RA, DEC, $P$, $\dot{P}$, binary orbit 
  parameters, pulse arrival time and absolute pulse phase for a given epoch, pulsar proper motion 
  etc.) are known with very high accuracy. Indeed, pulsar timing has reached the $10^{-15}$ fractional 
  level, which is comparable with the accuracy of atomic clocks. This is an essential requirement 
  for using these celestial objects as navigation beacons, as it enables one to predict the pulse 
  arrival time of a pulsar for any location in the solar system and beyond.    
 
\section{Principles of Pulsar-Based Navigation\label{PrinciplesPulsarNavigation}}

\begin{figure}[tb]
 \centerline{\includegraphics[width=7.75cm, angle=0,clip=]{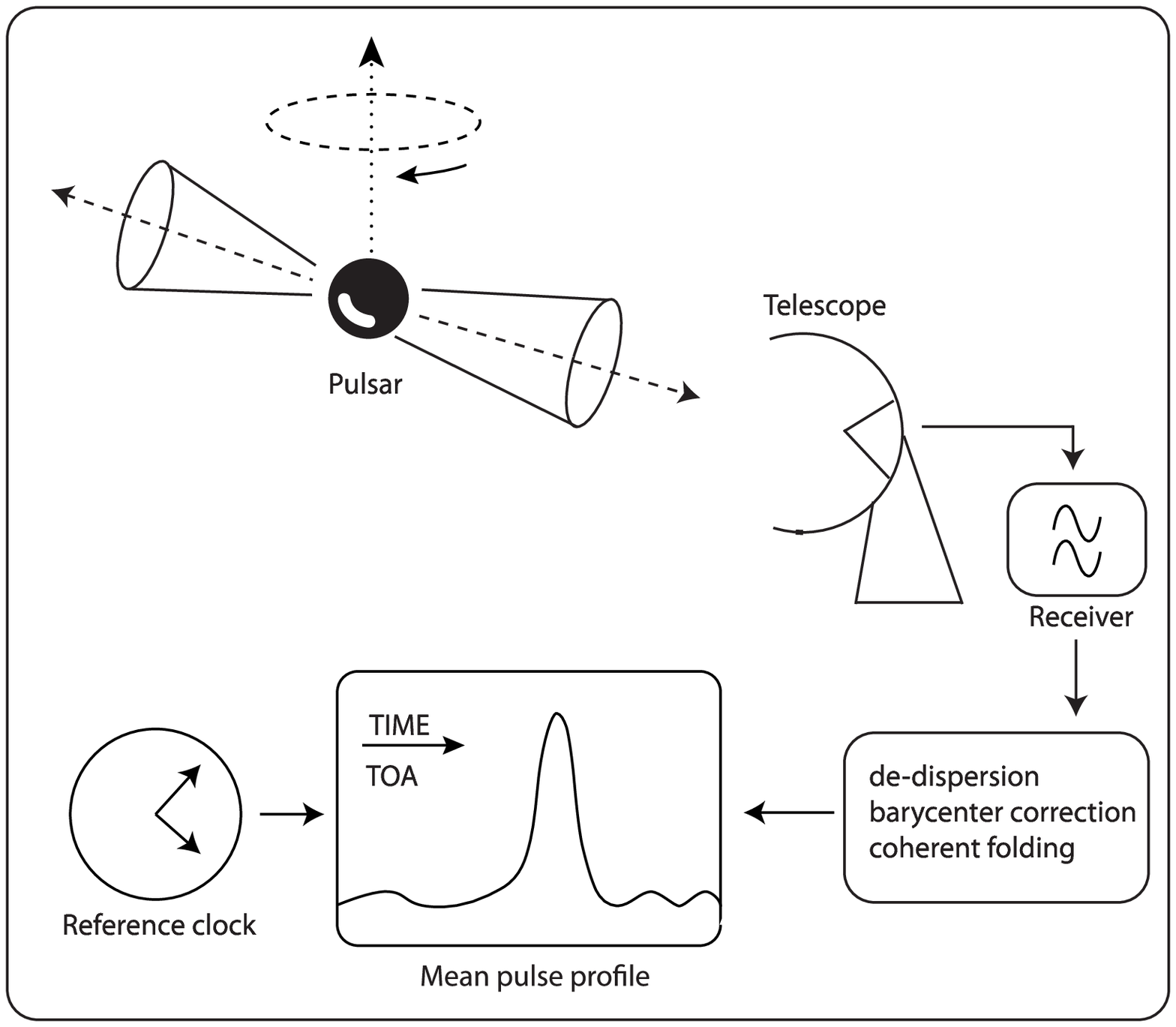}}
 \caption{\small Typical pulsar detection chain. The pulsar beams sweep across the radio antenna.
  Radio signals are recorded and analyzed in order to produce a mean pulse profile. The data 
  processing comprises a removal of dispersion effects caused by the interstellar medium 
  (``de-dispersion''), correction for the position and proper motion of the observatory 
  (``barycenter correction'') and coherent folding of many  pulses. The time of arrival (TOA) 
  of the pulse peak is measured against a reference clock.}\label{RadioDetectionChain}
 \end{figure}

\begin{figure}[t]
 \centerline{\includegraphics[width=7.75cm, angle=0,clip=]{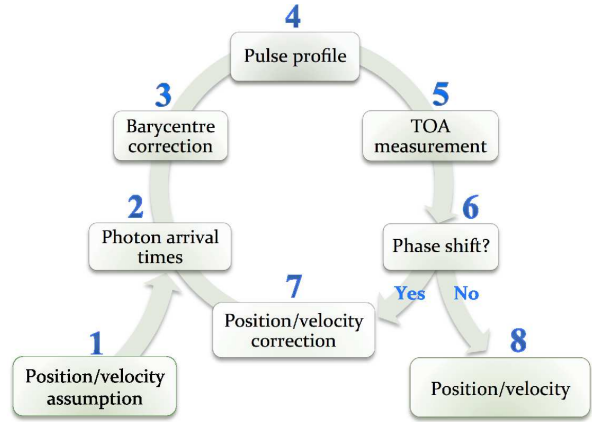}}
 \caption{\small Iterative determination of position and velocity by a pulsar-based
                 navigation system.}\label{image:iteration_loop}
 \end{figure}

  The concept of using pulsars as navigational aids is based on measurements of pulse arrival 
  times and comparison with predicted arrival times at a given epoch and reference location. 
  A typical chain for detecting, e.g., radio signals from a rotation-powered pulsar is shown in 
  Figure~\ref{RadioDetectionChain}. 
 An important step in this measurement is the barycenter correction of the observed photon 
 arrival times.  The pulsar ephemerides along with the position and velocity of the observer  
 are parameters of this correction. Using a spacecraft position that deviates from the true 
 position during the observation results in a phase shift of the pulse peak (or equivalently 
 in a difference in the pulse arrival time). Therefore, the position and velocity of the 
 spacecraft can be adjusted in an iterative process until the pulse arrival time matches with the 
 expected one. The corresponding iteration chain is shown in Figure~\ref{image:iteration_loop}.

\begin{figure}[t!!!!!!!]
 \centerline{\includegraphics[width=7.75cm, angle=0,clip=]{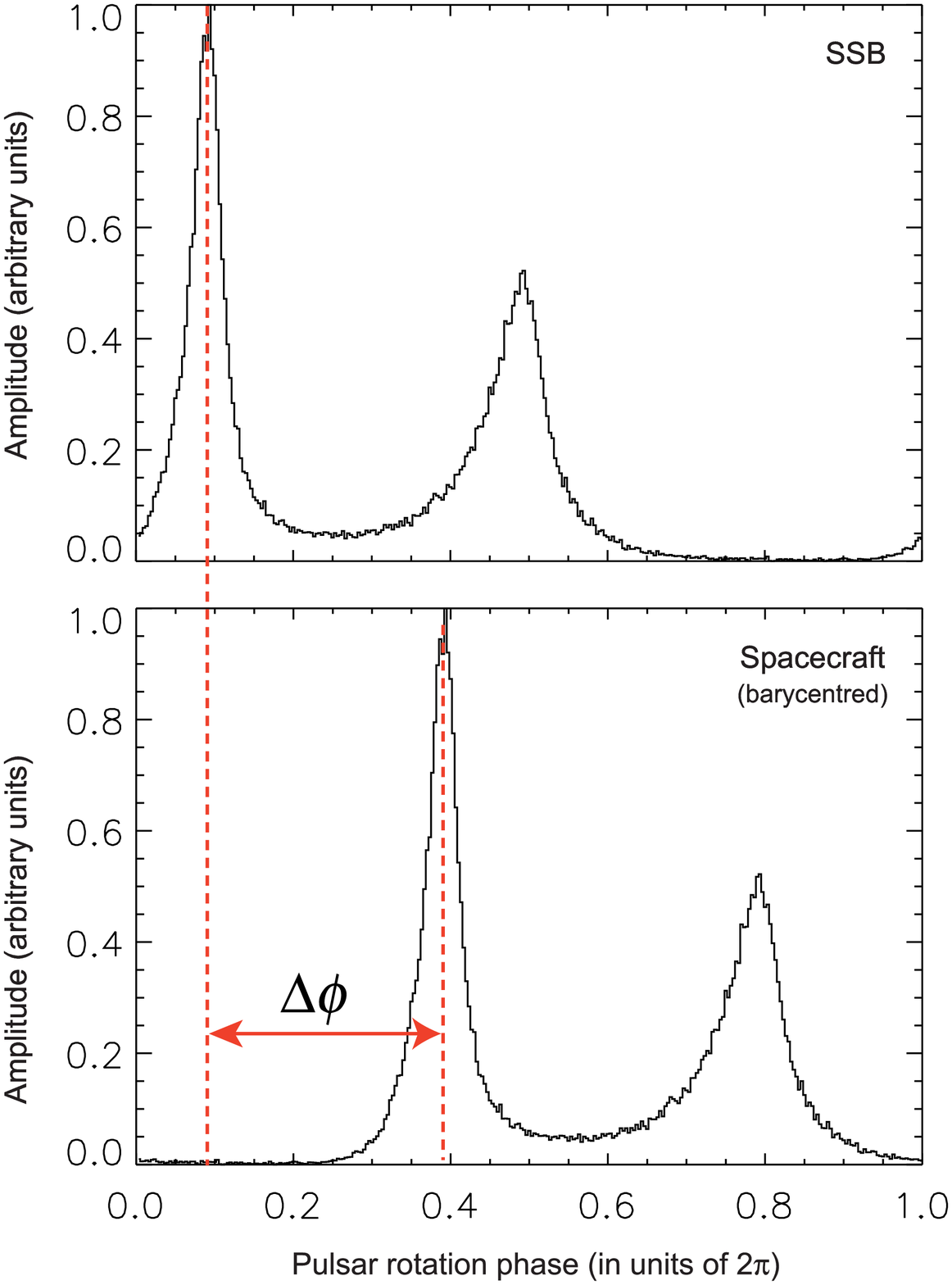}}
 \caption{\small Measuring the phase difference between the expected
  and measured pulse peak at an inertial reference location; e.g., the
  solar system barycenter (SSB). The top profile shows the main pulse peak
  location as expected at the SSB. The bottom profile is the one which has been
  measured at the spacecraft and transformed to the SSB by {\rm assuming} the
  spacecraft position and velocity during the observation. If the position and
  velocity assumption was wrong, a phase shift $\Delta \phi$ is observed.}
  \label{image:Crab_pulseprofile}
 \end{figure}

  An initial assumption of position and velocity is given by the planned orbit parameters
  of the spacecraft (1). The iteration starts with a pulsar observation, during which the 
  arrival times of individual photons are recorded (2). The photon 
 arrival times have to be corrected for the proper motion of the spacecraft by 
 transforming the arrival times (3) to an inertial reference location; e.g., the solar 
 system barycenter (SSB). This correction requires knowledge of the (assumed or deduced) 
 spacecraft position and velocity as input parameters. The barycenter corrected photon 
 arrival times allow then the construction of a pulse profile or pulse phase histogram 
 (4) representing the temporal emission characteristics and timing signature of the pulsar. 
 This pulse profile, which is continuously improving in significance during an observation, 
 is permanently correlated with a pulse profile template in order to increase the accuracy 
 of the absolute pulse-phase measurement (5), or equivalently, pulse arrival time (TOA).  
 From the pulsar ephemeris that includes the information of the absolute pulse phase for 
 a given epoch, the phase difference $\Delta\phi$ between the measured and predicted pulse 
 phase can be determined (cf.~Figure~\ref{image:Crab_pulseprofile}). 
 In this scheme, a phase shift (6) with respect to the absolute pulse phase corresponds to a 
 range difference $\Delta x = c P (\Delta\phi + n)$ along the line of sight toward the observed 
 pulsar. Here, $c$ is the speed of light, $P$ the pulse period, $\Delta\phi$ the phase shift and 
 $n=0, \pm 1, \pm 2,\ldots$ an integer that takes into account the periodicity of the observed pulses. 
 If the phase 
 shift is non-zero, the position and velocity of the spacecraft needs to be corrected accordingly 
 and the next iteration step is taken (7). If the phase shift is zero, or falls below a certain 
 threshold, the position and velocity used during the barycenter correction was correct (8) and 
 corresponds to the actual orbit of the spacecraft. 

 \begin{figure}[t!!!!!!!]
 \centerline{\includegraphics[width=7.75cm, angle=0,clip=0]{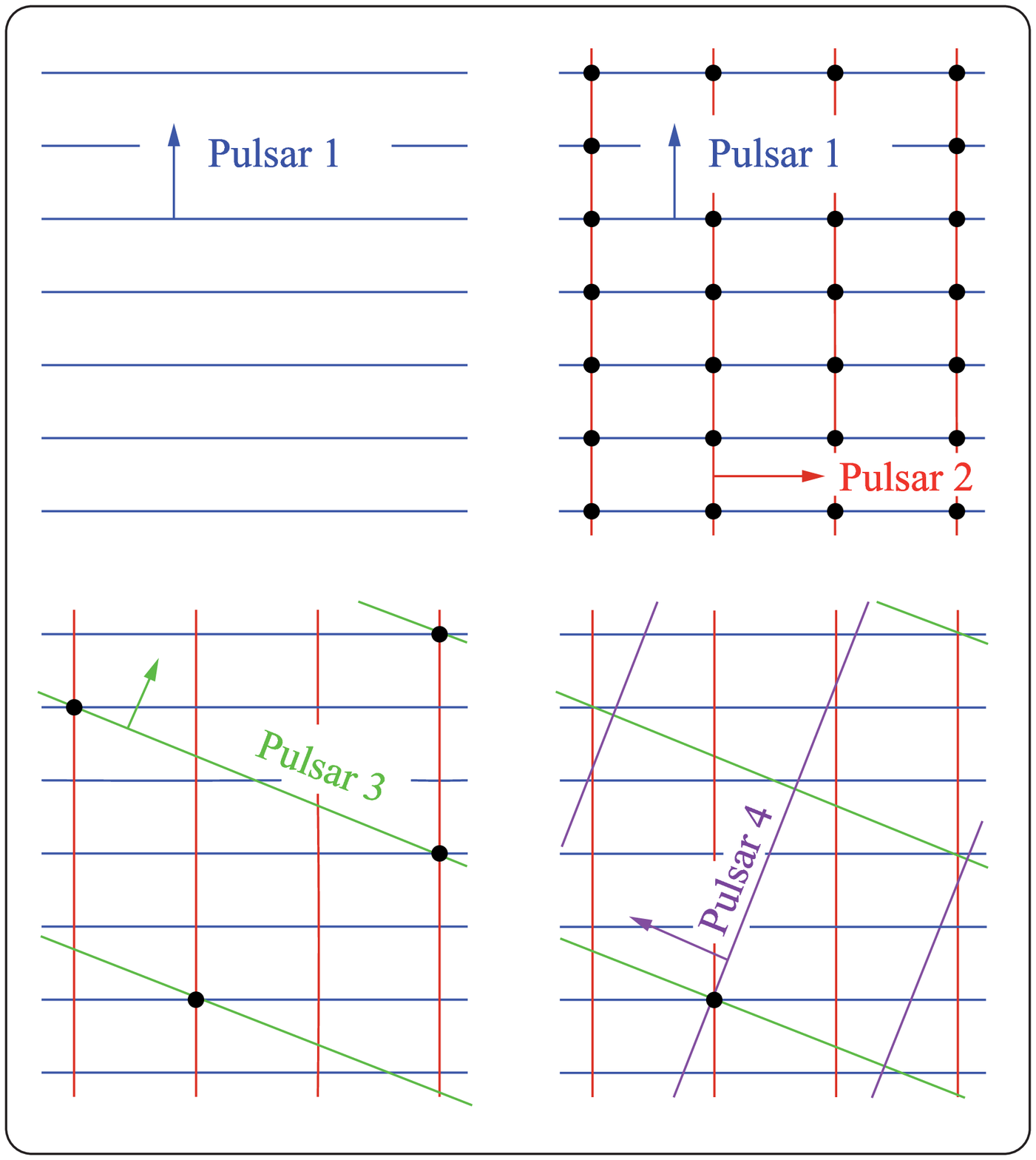}}
 \caption{\small Solving the ambiguity problem by observing four pulsars (drawn in two dimensions).   
  The arrows point along the pulsar's lines-of-sight. Straight lines represent planes of constant     
  pulse phase; black dots indicate intersections of planes.}\label{image:three_pulsars}               
 \end{figure} 

 A three-dimensional position fix can be derived from observations of at least three  
 different pulsars (cf.~Figure~\ref{image:three_pulsars}). If on-board clock calibration 
 is necessary, the observation of a fourth pulsar is required.

 Since the position of the spacecraft is deduced from the phase (or pulse arrival time)
 of a periodic signal, ambiguous solutions may occur. This problem can be solved by constraining
 the domain of possible solutions to a finite volume around an initial assumed position 
 \citep{bernhardt2010, bernhardt2011}, or by observing  additional pulsars as illustrated 
 in Figure~\ref{image:three_pulsars}. 

\section{Radio Antenna for Pulsar-Based Navigation}\label{RadioAntenna} 

 Pulsars emit broadband electromagnetic radiation. Therefore, the observing device of 
 a pulsar-based navigator can be optimized for the waveband according to the highest number 
 of bits per telescope collecting area, power consumption, navigator weight, cost and compactness. 
 Especially for the
 question of navigator compactness it is important to estimate what size a radio
 antenna would have to have in order to detect the emission from pulsars in a 
 reasonable integration time. In order to estimate this we assume pulsar parameters that are 
 typical for millisecond pulsars. As the radio flux from 
 pulsars shows a $\nu^{-1.5}$ dependence, observations at lower frequencies seem to 
 be preferred, but scintillation and scattering effects are stronger at lower 
 frequency. For a navigation system operating in the radio band of the electromagnetic spectrum, 
 the L-band at 21\,cm might therefore be best suited.

 For a pulsar detection we require a signal-to-noise ratio of $S/N=10$, a minimum 
 integration time of $t_\text{int}=3600$\,s and assume a frequency bandwidth of $\Delta\nu=100$\,MHz.  
 For the receiver noise temperature we take $T_\text{rec}=100$\,K. A lower temperature would 
 require active, e.g., cryogenic cooling, which would increase cost, weight, and power 
 consumption of a navigator. Furthermore, active cooling would severely limit the lifetime of the navigator 
 due to consumables like helium. For the sky temperature
 we take $T_\text{sky}=5$\,K and for the telescope efficiency $\epsilon=0.5$. If $A_\text{ant}$ 
 is the geometrical antenna area, the effective antenna area computes as 
 $A_\text{eff}=\epsilon\, A_\text{ant}$. For the period of the pulsar we assume $P=10$\,ms
 and for the pulse width $W=2$\,ms. For the average flux density we adopt 
 $\Delta S=10$\,mJy. Using the canonical sensitivity equation \citep{lorimer+kramer2005} 
 which corresponds to the radiometer equation applied to pulsar observations
 \begin{align*}
  \Delta S_\text{min} = \frac{2k}{\epsilon A_\text{ant}} 
                        \frac{(T_\text{rec}+T_\text{sky})}{\sqrt{n_p\, t_\text{int}\, \Delta\nu}}
                        \sqrt{\frac{W}{P-W}}
 \end{align*}
 and converting it to the geometrical antenna area, including the $S/N$ requirement, we get:
 \begin{align*}
   A_\text{ant} = \frac{S}{N} 
                  \frac{2k}{\epsilon \Delta S} 
                  \frac{(T_\text{rec}+T_\text{sky})}{\sqrt{2\,t_\text{int}\, \Delta\nu}}
                  \sqrt{\frac{W}{P-W}}  
 \end{align*}
 Here we have also assumed that both polarizations are averaged ($n_p=2$). From this we 
 compute an antenna area of $A_\text{ant}\approx 342$\,m$^2$ for the parameters specified above.
 For a parabolic antenna it would mean a radius of about $\sqrt{A_\text{ant} / \pi}\approx 
 10.5$\,m. Increasing the integration time to 4 hours, we find an area of 
  $\approx 171$\,m$^2$, which corresponds to a radius of $\approx 7.3$\,m. For comparison,  
 the radius of the communication antenna used on Cassini 
 and Voyager is 2\,m. It may depend on the satellite platform what size and antenna weight
 is acceptable, but a parabolic antenna does not seem to be very practical for navigation purposes.
 A navigator would have to observe several pulsars, either at the same time 
 or in series. The pulsars must be located in different sky regions in order to get an 
 accurate navigation result in the $x$, $y$ and $z$ direction.  This, however, means that 
 one either rotates the parabolic antenna or, alternatively, the whole satellite  to get the
 pulsar signals into the antenna focus. Rotating the satellite, though, would mean that 
 the communication antenna will not point to earth any more, which is undesirable.
 On earth satellites and space missions to the inner solar system, 
 power is usually generated by solar panels. Rotating the satellite would then also 
 mean to bring the solar panels 
 out of optimal alignment with the sun, which is another counter argument for using a 
 parabolic antenna, not to mention the effects of shadowing of the solar panels.

 It thus seems more reasonable to use dipole-array antennas for the pulsar observations.
 Single dipole antennas organized, e.g., as antenna patches 
 could be used to build a larger phased array antenna. Such a phased array would still 
 be large and heavy, though. Depending on the frequency, $10^4$--$10^5$ single patches 
 are required. There have been no phased-array antennas of that size been build for use 
 in space so far, although smaller prototypes exist \citep{datashvili2005}. From them one may 
 estimate the weight of such antenna arrays. Assuming an antenna thickness of 1\,cm
 and an averaged density of the antenna material of 0.1\,g/cm$^3$ still yields an antenna 
 weight of 170\,kg for the 170\,m$^2$ patched antenna array. The signals from the single 
 dipole antennas have to be correlated in phase to each other, which means that all patches 
 have to be connected to each other by a wired mesh and phase correlators. If this phase 
 correlation and the real-time coherent correction for the pulse broadening by 
 interstellar dispersion is done by software, it requires a computer with a Terra-flop GPU of 
 about 500\,W power consumption. A clear advantage of a phased antenna array would be 
 that it allows to observe different pulsars located in different sky regions at the
 same time. That means of course that such an antenna can be smaller if the same $S/N$
 ratio is to be achieved for a number $N_\text{sources}$  of sources within a given time. 
 With a single-dish antenna one would have to increase its diameter by $N_\text{sources}^{1/4}$ 
 if these had to be observed within the same given time interval.

\section{Using X-ray Signals from Pulsars for Spacecraft Navigation\label{XNAV}}

 The increasing sensitivity of the X-ray observatories ROSAT, RXTE, XMM-Newton and Chandra allowed 
 for the first time to explore in detail the X-ray emission properties of a larger sample of rotation-powered  
 pulsars. The discovery of pulsed X-ray emission from millisecond pulsars 
 \citep{becker+truemper1993}, the determination of the X-ray efficiency of rotation-powered pulsars 
 \citep{becker+truemper1997} as well as discoveries of X-ray emission from various pulsars 
 \citep[e.g.,][]{becker+truemper1999, becker2005, becker2006} and their detailed spatial, spectral 
 and timing studies are just a few of many accomplishments worth mentioning in this context 
 \citep[e.g.,][]{pavlov2001, kuiper2002, weisskopf2004, DeLuca2005, beckerBook2009, hermsen2013}. 
 With these new results at hand it was only natural to start looking at their applicability to, e.g., 
 spacecraft navigation based on X-ray data from pulsars. The prospects of this application are 
 of even further interest considering that low-mass X-ray mirrors, which are an important requirement 
 for a realistic implementation of such a navigation system, have been developed for future X-ray observatories.

 Given the observational and systematical limitations mentioned above it was a question 
 of general interest, which we found not considered with sufficient gravity in the literature, 
 whether it would be feasible to navigate a spaceship on arbitrary orbits by observing X-ray 
 pulsars. 

 In order to address the feasibility question we first determined the accuracy that can be achieved
 by a pulsar-based navigation system in view of the still limited information we have today 
 on pulse profiles and absolute pulse phases in the X-ray band. To overcome the 
 limitations introduced by improper fitting functions, undefined pivot points of pulse peaks 
 and phase shifts by an unmodeled energy dependence in the pulsed signal, we constructed pulse 
 profile templates for all pulsars for which pulsed X-ray emission is detected. Where supported 
 by photon statistics, templates were constructed for various energy ranges. These energy ranges 
 were chosen in order to optimize the $S/N$ ratio of the pulsed signal while sampling as much as 
 possible of the energy dependence of the X-ray pulses.

 In the literature various authors applied different analysis methods and often used different 
 definitions for pulsed fraction and pulse peak pivot points. Reanalyzing all data from X-ray
 pulsars available in the public XMM-Newton, Chandra and RXTE data archives was therefore a 
 requirement to reduce systematic uncertainties that would have been introduced otherwise. 
 The result is a database containing the energy dependent X-ray pulse profiles, 
 templates and relevant timing and spectral properties of all X-ray pulsars that have been
 detected so far \citep{prinz2010, breithuth2012}. 

 According to the harmonic content of an X-ray pulse the templates were obtained by fitting 
 the observed pulse profiles by series of Gaussian and sinusoidal functions. The database  
 further includes information on the local environment of a pulsar, i.e., whether it is surrounded 
 by a plerion, supernova remnant or whether it is located in a crowded sky region like a globular 
 cluster. The latter has a severe impact on the detectability of the X-ray pulses as it reduces
 the $S/N$ ratio of the pulsed emission by the DC emission from background sources. This in turn
 is important for the selection of the optimal pulsars that emit pulses, e.g., in the hard 
 band (above $\approx 3$\,keV) in order to blend away the softer emission from a supernova remnant or plerion. 

 The pulse profile templates allow us to measure pulse arrival times with high accuracy even  
 for sparse photon statistics by using a least-square fit of an adequately adjusted template. 
 The error of pulse-arrival-time measurements is dominated by the 
 systematic uncertainty that comes with the limited temporal resolution of the observed pulse 
 profiles used to construct the templates. The statistical error in fitting a measured profile
 by a template was found to be much smaller in all cases. Assuming that the temporal resolution  
 of the detector is not the limiting factor, the temporal resolution of a pulse profile is 
 given by the widths of the phase bins used to represent the observed X-ray pulse. The bin 
 width, or the number of phase bins applied, is a compromise between maximizing the $S/N$ ratio 
 per phase bin while sampling as much of the harmonic content as possible. Denoting the 
 Fourier-power of the $i$-th harmonic by $R_i$ and taking $m$ as the optimal number of harmonics 
 deduced from the H-test \citep{DeJaeger1989}, an exact expression for the optimal number of phase 
 bins is given by $ M = 2.36 (\sum_{i=1}^{m} i^2 R_i^2)^{1/3}$ \citep{becker+truemper1999}. This
 formula compromises between information lost due to binning (i.e., zero bin width to get all 
 information), and the effect of fluctuations due to finite statistics per bin (i.e., bin width 
 as large as possible to reduce the statistical error per bin). The total error (bias plus 
 variance) is minimized at a bin width of $1/M$. We applied this in the reanalysis of pulse profiles 
 in our database. Pulsed fractions were computed by applying a bootstrap method 
 \citep{becker+truemper1999}, which again leads to results that are not biased by the 
 observers ``taste'' on where to assume the DC level in a profile. 

 The minimal systematic phase uncertainty for the pulse profile templates in our database is 
 of the order of 0.001 \citep{bernhardt2010}. This uncertainty multiplied by the rotation period 
 $P$ of the pulsar
 yields the uncertainty in pulse arrival time due to the limited information we have 
 on the exact X-ray pulse profile. Multiplying this in turn by the speed of light yields the 
 spacecraft's position error along the line of sight to the pulsar. It is evident by 
 the linear dependence on $P$ that millisecond pulsars are better suited for navigation than those 
 with larger rotation periods.
 
 The precision of a pulsar-based navigation system thus strongly depends on the choice of 
 pulsars and the accuracy of pulse arrival measurements, which is subject to the quality 
 of the available templates, accuracy of the on-board clock and clock calibration. As mentioned 
 above, in order to obtain three-dimensional position information, timing of at least three 
 different pulsars has to be performed. The spatial arrangement of these pulsars is another
 parameter of the achievable accuracy. Our simulations show that the systematic error of position determination
 can be reduced significantly by choosing a pulsar triple that is optimal in the sense that 
 the pulsars are nearly perpendicular to each other. 
 Since a pulsar might be obscured by the sun or a planet and, therefore, its availability for navigation 
 depends on the current position of the spacecraft, the optimal pulsar triple has
 to be selected from a ranking of possible pulsar combinations. The following Table \ref{table:accuracy}
 represents the ranking of pulsar combinations, which according to our analysis provide the highest 
 position accuracy via pulsar navigation (Bernhardt et al.~2010). 

\begin{table}[!h]
\footnotesize
\centering                
\begin{tabular}{c r r c} \hline\hline\\[-2ex]
 Rank & \multicolumn{3}{c}{Pulsar 3-Combination}\\\hline\\[-2ex]
  1  &  B1937$+$21   &	B1821$-$24    &		J0030$+$0451 \\
  2  &	B1937$+$21   &	B1821$-$24    &		J1023$+$0038 \\
  3  &	B1821$-$24   &	J0030$+$0451  & 	J0437$-$4715\\
  4  &	B1937$+$21   &	J1023$+$0038  & 	J0218$+$4232 \\
  5  &	B1821$-$24   &	J1023$+$0038  & 	J0437$-$4715 \\
  6  &	B1937$+$21   &	J0030$+$0451  & 	J0218$+$4232 \\
  7  &	B1937$+$21   &	B1821$-$24    &		J0437$-$4715 \\
  8  &	B1937$+$21   &	J0218$+$4232  & 	J0437$-$4715 \\
  9  &	B1821$-$24   &	J0218$+$4232  & 	J0437$-$4715 \\
 10  &	J1023$+$0038 & 	J0218$+$4232  & 	J0437$-$4715 \\ \\[-2ex] \hline\hline
\end{tabular}
\caption{\small Ranking of pulsar 3-combinations according to the position accuracy achievable  
 when using them in a navigation system based onf X-ray pulsars . All listed sources are solitary 
 millisecond pulsars except J0437$-$4715, which is in a binary.} 
 \label{table:accuracy}
\end{table}

\begin{figure}[h!!!!!!!!!!!!!!!!!] 
 \centerline{\includegraphics[width=7.75cm, angle=0,clip=0]{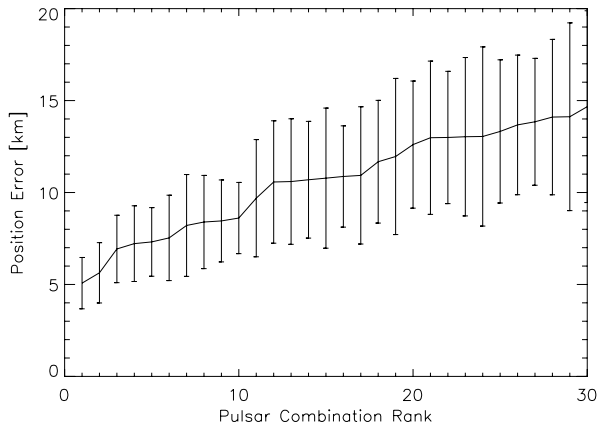}}   
  \caption{\small Spacecraft position error as a function of possible pulsar 3-combinations. 
  The diagram shows the mean position errors and standard deviations for the best 30 combinations.
  From \citet{bernhardt2010}.} \label{image:accuracy}
 \end{figure}

 For the pulsars ranked highest in Table \ref{table:accuracy} we found position errors 
 of about 5 km as a lower limit (cf.~Figure~\ref{image:accuracy}). The ranking is independent 
 from a specific spacecraft orbit, but was obtained under the assumption that the navigation system is capable
 of measuring pulse profiles with the same level of detail and accuracy as the ones used in the simulation. Indeed, this 
 is a severe limitation as those pulse profiles where obtained by powerful X-ray observatories 
 like XMM-Newton and Chandra. It is unlikely, due to weight constraints and power 
 limitations, that a navigation system will have similar capacities in terms of collecting power, 
 temporal resolution and angular resolution.

 An improved accuracy can be achieved by means of pulse profile templates of better quality.
 This, in turn, calls for deeper pulsar observations by XMM-Newton
 or Chandra as long as these observatories are still available for the scientific community. It would be 
 a valuable task worthwhile the observing time, especially as it is unclear what missions will follow 
 these great observatories and whether they will provide detectors with sufficient 
 temporal resolution and on-board clock accuracy. 

 We extended our simulations in order to constrain the technological parameters 
 of possible navigation systems based on X-ray pulsars.
 The result of this ongoing project will be a high-level design of a pulsar navigator
 that accounts for boundary conditions (e.g., weight, cost, complexity and power consumption) 
 set by the requirements of a specific spacecraft and mission design.

\begin{figure*}[p]  
 \centerline{\includegraphics[width=13.2cm, angle=0,clip=0]{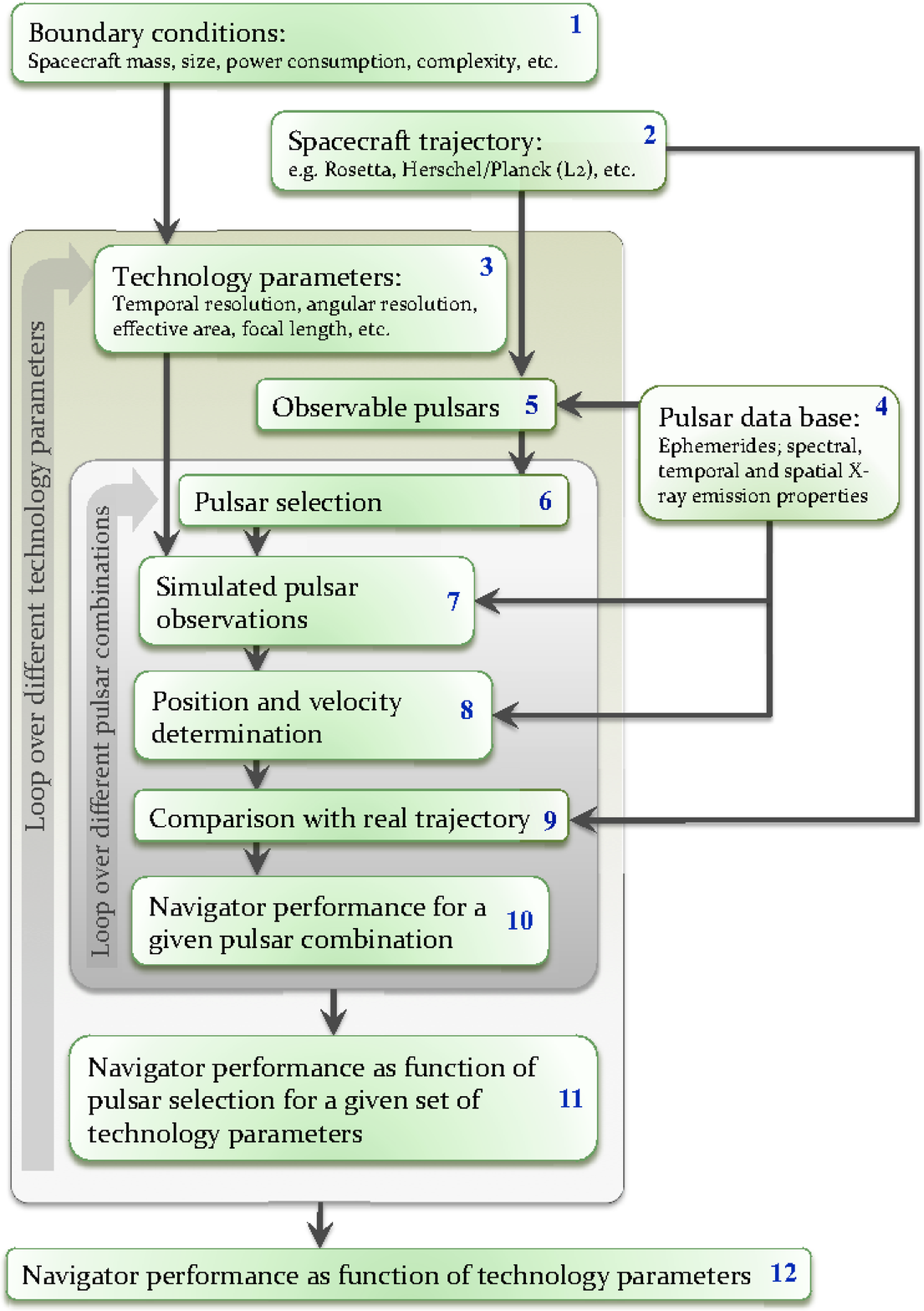}}  
  \caption{\small Work flow and logic of the simulations performed for a       
  technology requirement study and demonstrator high-level design of a pulsar-based navigator.}              
  \label{image:simulations}
 \end{figure*}

 The chart shown in Figure~\ref{image:simulations} illustrates the work logic of our current 
 simulations. Navigator boundary conditions (1) and spacecraft orbit (2) are predefined and constrain 
 the technology parameters (3) of the navigator's X-ray detector, mirror system and on-board electronics.
 Examples of parameters that will be analyzed in the simulations are detector technology, 
 temporal and energy resolution, on-board-clock accuracy and stability, mirror technology 
 along with collecting power, angular and spatial resolution, focal length, field of view -- just to 
 mention the most important ones.
 Given their X-ray emission properties some pulsars may not be detectable by the navigator 
 because of its limited angular resolution and sensitivity. The properties of the detector and 
 mirror system along with the trajectory of the 
 virtual navigator thus determine the set of available pulsars (5), from which a 
 suitable selection has to be made (6) according to a predefined ranking of pulsar triples. 
 Using an X-ray-sky simulator (7), which was developed to simulate observations of the future 
 X-ray observatories eROSITA and IXO and which we have modified to include the temporal 
 emission properties of X-ray pulsars, we will be able to create X-ray datasets in FITS-format with 
 temporal, spatial and energy information for the pulsars observed by our virtual navigator.
 The simulated event files have the same standard FITS-format as those of XMM-Newton and/or Chandra, so 
 that standard software can be used to analyze these data. An autonomous data reduction will then 
 perform the data analysis and TOA measurement in order to obtain the position and velocity of the virtual observer
 (8). Correlating the result with the input trajectory (9) yields the accuracy of the simulated measurements 
 for a given spacecraft orbit (10), and hence the overall performance of the navigator 
 as a function of the specified detector and mirror system (11,12).

\section{X-ray Detector and Mirror Technology for Pulsar-Based Navigation}\label{XNAV_hardware}
 
 The design of an X-ray telescope suitable for navigation by X-ray pulsars will be a compromise 
 between angular resolution, collecting area and weight of the system. The currently operating 
 X-ray observatories XMM-Newton and Chandra have huge collecting areas of 0.43 m$^2$ and 
 0.08 m$^2$ (at 1 keV), respectively \citep[and references therein]{friedrich2008} and, in the  
 case of Chandra, attain very good angular resolution of less than 1 arcsecond. Their 
 focusing optics and support structures, however, are very heavy. To use their mirror
 technology would be a show-stopper for a navigation system. 
   
 In recent years, ESA and NASA have put tremendous effort into the development of low-mass 
 X-ray mirrors, which can be used as basic technology for future large X-ray observatories 
 and small planetary exploration missions. Table \ref{table:mirrorweight} summarizes the angular  
 resolution and mass of X-ray mirrors used for XMM-Newton and Chandra as well as developed 
 for future X-ray missions. The light-weighted mirrors are of special interest for an X-ray 
 pulsar-based navigator. 

\begin{table}[h!!!!!!!!!!]
\footnotesize
\centering                
\begin{tabular}{l l l} \hline\hline\\[-2ex]
     &  Angular     & Mass per effective  \\
     &  resolution  & area (at 1 keV) \\\hline\\[-2ex]
   Chandra    & $0.5''$ & $18\,500$ kg/m$^2$ \\
   XMM-Newton & $14''$  & $2300$ kg/m$^2$ \\
   Silicon Pore Optics & $5''$   & $200$ kg/m$^2$ \\
   Glass Micropore Optics  & $30''$  & $25$ kg/m$^2$ \\ \\[-2ex] \hline\hline
\end{tabular}
 \caption[Table]{\small Comparison of current and future X-ray-mirror optics. From \citet{bavdaz2010}.} 
 \label{table:mirrorweight}
\end{table}

 A typical high-resolution X-ray telescope uses focusing optics based on the 
 Wolter-I design \citep{wolter1952}. The incoming X-ray photons are reflected under 
 small angles of incidence in order not to be absorbed and are focused by double 
 reflection off a parabolic and then a hyperbolic surface. This geometry allows for  
 nesting several concentric mirror shells into each other in order to enlarge the 
 collecting area and thereby improve the signal-to-noise ratio. A novel approach to 
 X-ray optics is the use of pore structures in a Wolter-I configuration \citep{bavdaz2003, bavdaz2010, beijersbergen2004a}. X-ray photons that enter 
 a pore are focused by reflections on the walls inside the pore. In contrast to 
 traditional X-ray optics with separate mirror shells that are mounted to a support 
 structure, pore optics form a monolithic, self-supporting structure that is lightweight, 
 but also very stiff and contains many reflecting surfaces in a compact assembly. Two 
 different types of pore optics have been developed, based on silicon and glass. 

 \begin{figure}[t!!!!!!!!!!!!!!!!!]
 \centerline{\includegraphics[width=7cm, angle=0,clip=0]{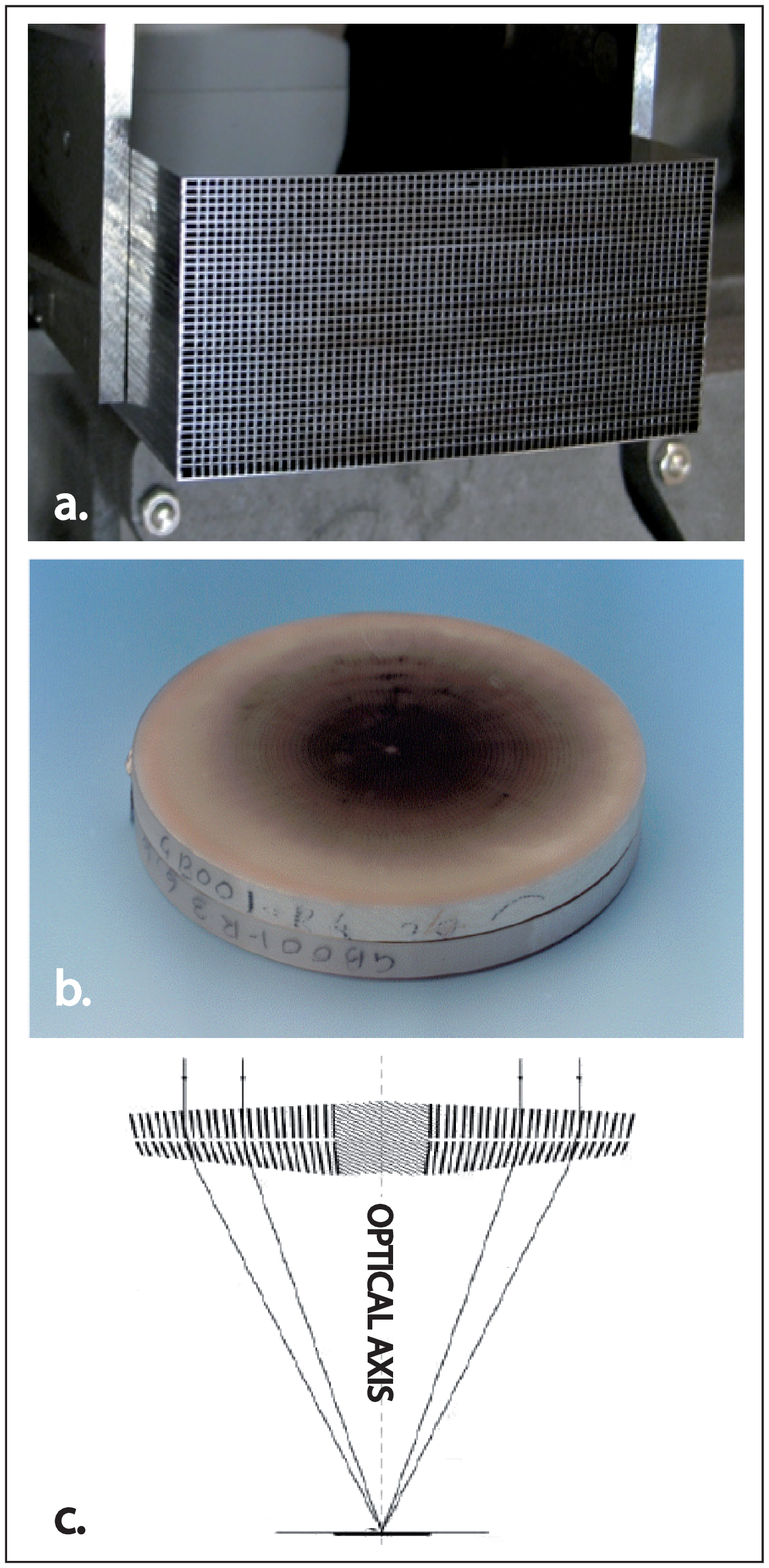}}              
  \caption[Table]{\small Silicon pore optics (a) and glass micropore optics (b)     
  represent novel developments for light-weighted X-ray mirrors of the next generation of X-ray 
  observatories. Both mirror types will be used in Wolter-I configuration (c) 
  to focus X-rays in a double reflection. Images from \citet{bavdaz2003, bavdaz2004, bavdaz2010}.}
  \label{image:novelmirror}
 \end{figure}

 \begin{itemize}
 \item	{\bf Silicon Pore Optics} 
         \citep{collon2009, collon2010, ackermann2009} use commercially available and mass 
         produced silicon wafers (Figure~\ref{image:novelmirror}a) from the semiconductor industry. 
         These wafers have a surface roughness that is sufficiently low to meet the requirements of 
         X-ray optics. A chemo-mechanical treatment of a wafer results in a very thin 
         membrane with a highly polished surface on one side and thin ribs of very accurate 
         height on the other side. Several of these ribbed plates are elastically bent to 
         the geometry of a Wolter-I system, stacked together to form the pore structure and 
         finally integrated into mirror modules (Figure~\ref{image:novelmirror}a). 
         Silicon Pore Optics are intended to be used 
         on large X-ray observatories that require a small mass per collecting area (on the 
         order of 200 kg/m$^2$) and angular resolution of about 5 arcseconds or better.

 \item	{\bf Glass Micropore Optics} \citep{beijersbergen2004b, collon2007, wallace2007} 
         are made from polished glass blocks that are surrounded by a cladding glass with a lower 
         melting point. In order to obtain the high surface 
         quality required for X-ray optics, the blocks are stretched into small fibers, thereby 
         reducing the surface roughness. Several of 
         these fibers can be assembled and fused into multi-fiber bundles. Etching away the glass 
         fiber cores leads to the desired micropore structure, in which the remaining cladding glass 
         forms the pore walls (Figure~\ref{image:novelmirror}b). The Wolter-I geometry is reproduced by thermally slumping separate 
         multi-fiber plates. Glass Micropore Optics are even lighter than Silicon Pore Optics, but 
         achieve a moderate angular resolution of about 30 arcseconds. They are especially interesting 
         for small planetary exploration missions, but also for X-ray timing missions that require
         large collecting areas. The first implementation of Glass Micropore Optics on a flight program
         will be in the Mercury Imaging X-ray Spectrometer on the ESA/JAXA mission BepiColombo, 
         planned to launch in 2014 \citep{fraser2010}.

\end{itemize}

\begin{figure}[t!!!!!!!!!!!!!!!!!]
 \centerline{\includegraphics[width=7.75cm, angle=0,clip=0]{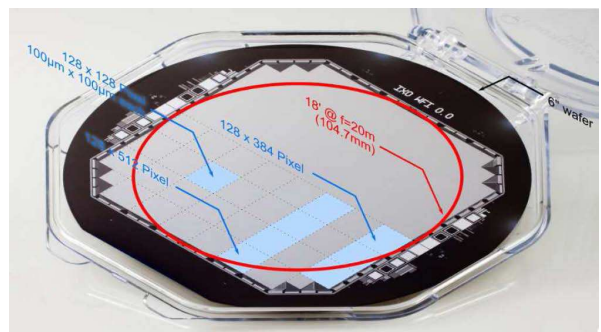}}              
  \caption[Table]{\small Mechanical sample of an Active Pixel (here 6-inch wafer-scale) 
  detector. Plotted over one hemisphere is the logical layout of the detector. It consists 
  of roughly $1024 \times 1024$ pixels of $100 \times 100$ \mbox{$\mu$m}$^2$ size. From \citet{lechner2010}.}
  \label{image:ActivePixelSensor}
 \end{figure}

\noindent
  Today's detector technology, as in use on XMM-Newton and Chandra, is not seen to provide 
  a detector design that is useful for a navigation system based on X-ray pulsars.
  Readout noise, limited imaging capability in timing-mode and out-of-time events invalidate 
  CCD-based X-ray detectors for application as X-ray-pulsar navigator.  Detectors 
  like those on RXTE that need gas for operation are also not suitable, given the limited live 
  time due to consumables. However, there are novel and promising detector developments performed 
  in semiconductor labs for the use in the next generation of X-ray observatories. 
  Two challenging examples, which are of potential interest for navigation, are:

 \begin{itemize}
 \item {\bf Silicon Drift Detectors (SDDs)} have only limited imaging capability but provide an energy resolution and are 
       capable of managing high counting rates of more than 10 Crab (1 Crab count rate $\approx 
       200\,000$\,cts/s). A detector based on this 
       technology was proposed for the High Time Resolution Spectrometer on IXO \citep{barret2010}. The detector technology itself has a high technical readiness. SDD-modules 
       developed in the Semiconductor Lab of the Max Planck Society are working already in the APXS 
       (Alpha Particle 
       X-ray Spectrometer) on-board NASA's Mars Exploration Rovers Spirit and Opportunity 
       and on the comet lander ROSETTA \citep{lechner2010}. Detectors based on an SDD 
       technology could be of use, e.g., in designing a navigator for a very specific orbit, 
       for which it is sufficient to navigate according to the signals of pulsars that emit
       their pulses in the hard X-ray band mostly so that the missing imaging capability 
       does not cause any restrictions on the $S/N$ ratio of the pulsed emission.

 \item {\bf Active Pixel Sensors (APS)} are an alternative and perhaps more flexible technology 
        (cf.~Figure~\ref{image:ActivePixelSensor}), which was 
        the proposed technology for the Wilde Field 
        Imager on IXO \citep{strueder2010} and the Low-Energy Detector (LED) on Simbol-X \citep{lechner2009}.
        This detector provides images in the energy band 0.1--25 keV, 
        simultaneously with spectrally and time resolved photon counting. The device, which 
        is under development in the MPE Semiconductor Laboratory, consists of an array of 
        DEPFET (Depleted p-channel FET) active pixels, which are integrated onto a common 
        silicon bulk. The DEPFET concept unifies the functionalities of both sensor and 
        amplifier in one device. It has a signal charge storage capability and is read out 
        demand. The DEPFET is used as unit cell of Active Pixel Sensors (APSs) with a scalable 
        pixel size from $50\,\mu$m to several mm and a column-parallel row-by-row readout with a 
        short signal processing time of $\le 4\,\mu$sec per row. As the pixels are individually 
        addressable the DEPFET APS offers flexible readout strategies from standard full-frame 
        mode to user-defined window mode. 

 \end{itemize}

\noindent
  The typical weight and power consumption of these detectors can be estimated from the 
  prototypes proposed for IXO and Simbol-X. The IXO Wide Field Imager with its 17 arcmin 
  field of view and $1024 \times 1024$ pixel design had an energy consumption 
  of $\le 22$ W. The Low Energy Detector on Simbol-X had $128 \times 128$ pixels and an energy 
  consumption of $\le 8$ W. Power consumption including electronics, filter wheel and 
  temperature control was ca.~250 W. The mass of the focal plane, including shielding 
  and thermal interface, was about 15 kg but could be reduced in a more specific design of 
  an X-ray-pulsar navigator.

\section{Concluding Remarks}
\begin{figure*}[t!!!!!!!!!!!!!!!!!]
 \centerline{\includegraphics[width=15.8cm, angle=0,clip=0]{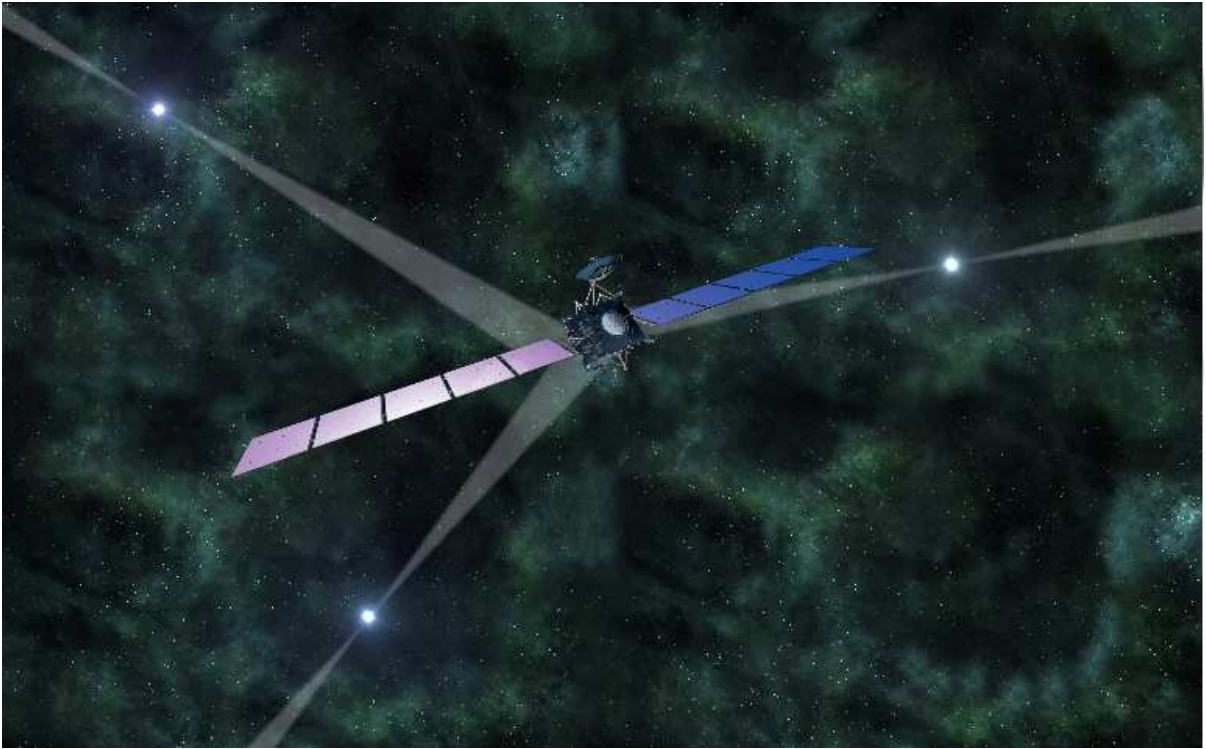}}
  \caption[Table]{\small Artist's impression of Rosetta, if it navigated in 
  deep space using pulsar signals. The characteristic time signatures of 
  pulsars are used as natural navigation beacons to determine the position 
  and velocity of the spacecraft.} \label{image:Rosetta}
 \end{figure*}

  The knowledge of how to use stars, planets and stellar constellations for navigation was
  fundamental for mankind in discovering new continents and subduing living space in ancient
  times. It is fascinating to see how history repeats itself in that a special population of stars
  may play again a fundamental role in the future of mankind by providing a reference for
  navigating their spaceships through the Universe (cf.~Figure \ref{image:Rosetta}).

  In the paper we have shown that autonomous spacecraft navigating  with pulsars is feasible
  when using either phased-array radio antennas of at least 150\,m$^2$ antenna area or
  compact light-weighted X-ray telescopes and detectors, which are currently developed for
  the next generation of X-ray observatories.

  Using the X-ray signals from millisecond pulsars we estimated that navigation would be
  possible with an accuracy of $\pm 5$\,km in the solar system and beyond. The error is dominated 
  by the inaccuracy of the pulse profiles templates that were used 
  for the pulse peak fittings and pulse-TOA measurements. As those are known with 
  much higher accuracy in the radio band, it is possible to increase the accuracy of  
  pulsar navigation down to the meter scale by using radio signals from pulsars for navigation.

  The disadvantage of radio observations in a navigator, though, is the large size 
  and mass of the phased-antenna array. As we saw in \S\,\ref{RadioAntenna}, the 
  antenna area is inversely proportional to the square root of the integration time;
  i.e., the same signal quality can be obtained with a reduced antenna size by increasing the observation time.
  However, the observing time is limited by the Allen variance of the receiving system and, therefore, 
  cannot become arbitrarily large. In addition,
  irradiation from the on-board electronics requires an efficient electromagnetic 
  shielding to prevent signal feedback. This shielding will further increase the 
  navigator weight in addition to the weight of the antenna.

  The optimal choice of the observing band depends on the boundary conditions given 
  by a specific mission. What power consumption and what navigator weight might be 
  allowed for may determine the choice for a specific wave band.

  In general, however, it is clear already today that this navigation technique
  will find its applications in future astronautics. The technique behind it is
  very simple and straightforward, and pulsars are available everywhere in the
  Galaxy. Today $\approx 2200$ pulsars are known. With the next generation of radio 
  observatories, like the SKA, it is expected to detect signals from about 
  20\,000 to 30\,000 pulsars \citep{smits2009}. 

 Finally, pulsar-based navigation systems can operate autonomously. This 
 is one of their most important advantages, and is interesting also for
 current space technologies; e.g., as augmentation of existing GPS/Galileo
 satellites. Future applications of this autonomous navigation technique might be 
 on planetary exploration missions and on manned missions to Mars or beyond.

\section*{Acknowledgments} 
 WB acknowledges discussion with David Champion (MPIfR), Horst Baier and Ulrich Walter (TUM).  
 MGB acknowledges support from and participation in the International Max Planck Research 
 School on Astrophysics at the Ludwig Maximilians University of Munich, Germany.

\bibliographystyle{apalike2} 
\bibliography{bibliography}

\end{document}